\documentclass[conference]{IEEEtran}
\IEEEoverridecommandlockouts
\usepackage{cite}
\usepackage{amsmath,amssymb,amsfonts}
\usepackage{algorithmic}
\usepackage[dvipdfmx]{graphicx}
\usepackage{textcomp}
\usepackage{xcolor}
\def\BibTeX{{\rm B\kern-.05em{\sc i\kern-.025em b}\kern-.08em
    T\kern-.1667em\lower.7ex\hbox{E}\kern-.125emX}}

\usepackage{multirow}
\usepackage{epstopdf}
\usepackage{comment}
\usepackage{subfigure}
\usepackage{colortbl}
\usepackage{arydshln}
\usepackage{here}
\usepackage{subfigure}
\usepackage{lscape}
\usepackage{booktabs}
\usepackage[utf8]{inputenc}
\usepackage{tabularx}
\usepackage{setspace}
\usepackage{mathtools}

\begin{document}

\title{Refining knowledge transfer on audio-image temporal agreement for audio-text cross retrieval
\thanks{This work was supported by JSPS KAKENHI Grant Numbers JP20H00613, 22H03639, and 23K16908.}
}

\author{
\IEEEauthorblockN{Shunsuke Tsubaki$^{1}$, Daisuke Niizumi$^{2}$, Daiki Takeuchi$^{2}$, Yasunori Ohishi$^{2}$, Noboru Harada$^{2}$, Keisuke Imoto$^{1}$ }
\IEEEauthorblockA{\textit{$^1$ Doshisha University, Japan, $^2$ NTT Corporation, Japan}}
}

\maketitle

\begin{abstract}
The aim of this research is to refine knowledge transfer on audio-image temporal agreement for audio-text cross retrieval.
To address the limited availability of paired non-speech audio-text data, learning methods for transferring the knowledge acquired from a large amount of paired audio-image data to shared audio-text representation have been investigated, suggesting the importance of how audio-image co-occurrence is learned.
Conventional approaches in audio-image learning assign a single image randomly selected from the corresponding video stream to the entire audio clip, assuming their co-occurrence. 
However, this method may not accurately capture the temporal agreement between the target audio and image because a single image can only represent a snapshot of a scene, though the target audio changes from moment to moment.
To address this problem, we propose two methods for audio and image matching that effectively capture the temporal information: (i) Nearest Match wherein an image is selected from multiple time frames based on similarity with audio, and (ii) Multiframe Match wherein audio and image pairs of multiple time frames are used.
Experimental results show that method (i) improves the audio-text retrieval performance by selecting the nearest image that aligns with the audio information and transferring the learned knowledge.
Conversely, method (ii) improves the performance of audio-image retrieval while not showing significant improvements in audio-text retrieval performance.
These results indicate that refining audio-image temporal agreement may contribute to better knowledge transfer to audio-text retrieval.
\end{abstract}


\section{Introduction}
\label{sec:intro}
\begin{figure}[t!]
  \centering
  \centerline{\includegraphics[width=\linewidth,pagebox=cropbox]{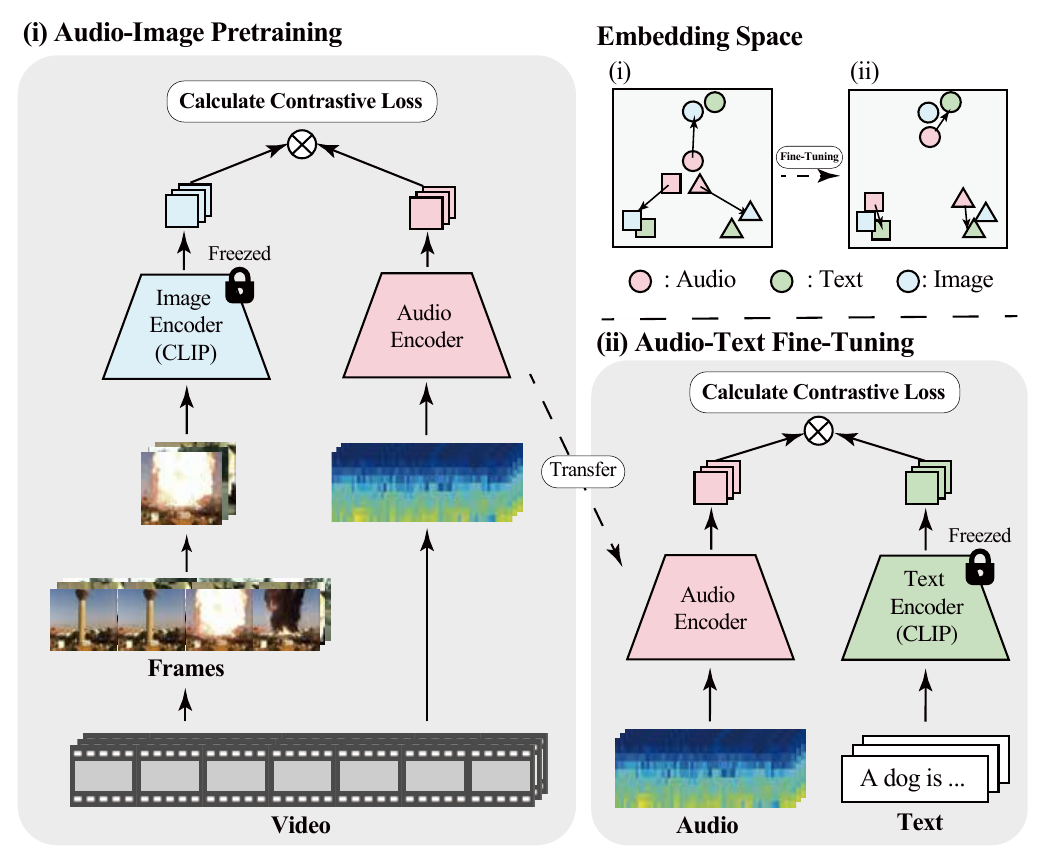}}
  \vspace{-8pt}
  \caption{Contrastive audio-image pretraining and audio-text fine-tuning and diagram of the embedding space. The image, audio, and text are colored blue, pink, and green, respectively.}
  \label{fig:multi}
  \vspace{-1pt}
\end{figure}
\begin{figure*}[t!]
  \centering
  \centerline{\includegraphics[width=\linewidth,pagebox=cropbox]{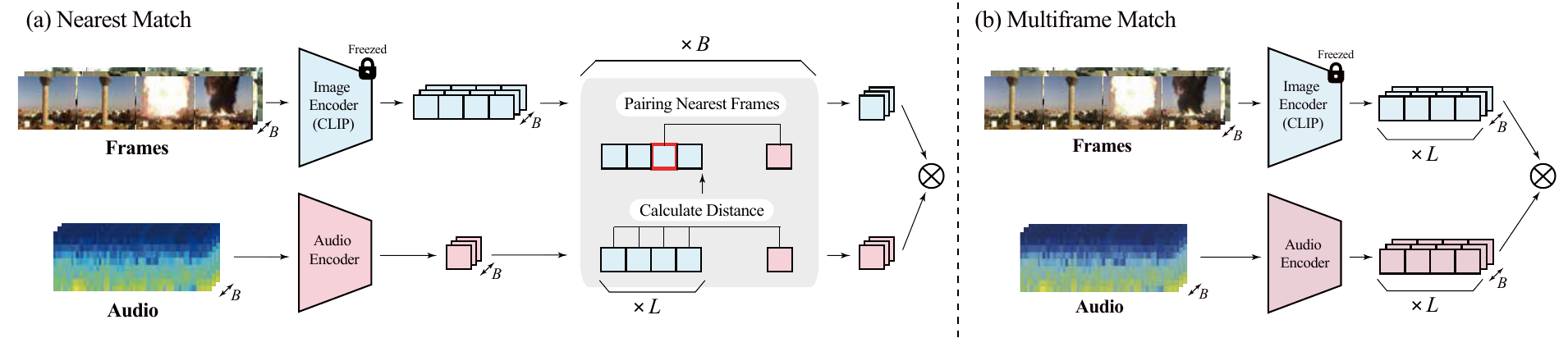}}
  \vspace{-7pt}
  \caption{Overview of our proposed audio-image pretraining scheme. On the left, the method is based on Nearest Match. On the right, the method is based on Multiframe Match. $B$ is the batch size during training.}
  \label{fig:proposed_model}
\end{figure*}

In recent years, there has been a growing interest in multimodal retrieval~\cite{MulRet_Kaiye2016}, which involves searching various combinations of modalities such as image-text~\cite{Vonam2019Imagetext,cao2022itsurvey}, video-text~\cite{fang2021clip2video, xu2021videoclip}, audio-text~\cite{guzhov2022audioclip,wu2022wav2clip}, and video-text-audio~\cite{ruan2023accommodating,akbari2021vatt}.
Among these modalities, audio-text retrieval has attracted attention because of its practical possibilities, such as improving accessibility of audio and audiovisual material for people with hearing impairments~\cite{TexAud_Oncescu2021}.
However, there are limitations regarding the availability of paired non-speech audio-text data compared with paired image-text data.
This limitation is primarily due to the difficulty of collecting such data by crawling, unlike, for example, the readily available Wikipedia-based Image Text Dataset (WIT)~\cite{Srinivasan2021WIT} which is image-text pair data.
For instance, while the Microsoft Common Objects in COntext (COCO) captions~\cite{chen2015cococap} have more than 1,500K image-text pairs, Clotho~\cite{drossos2020clotho} and AudioCaps~\cite{kim-NAACL-HLT-2019AudioCaps} have only about 35K and 46K audio-text pairs, respectively.
As a result, audio-text retrieval struggles to attain a level of performance that meets the requirements for practical applications~\cite{Koepke2022atretrievalque}.

To address this problem, researchers have explored methods of learning audio-text alignments using limited amounts of audio-text pairs~\cite{wu2022wav2clip,guzhov2022audioclip,vipant_zhao2022g} (see Fig.~\ref{fig:multi}).
These studies have exploited the large amount of paired image-text and audio-image data available online.
They involve positioning audio and images within an embedding space centered around the image modality, thereby implicitly capturing their relationships, as shown in Fig.~\ref{fig:multi} (i).
The Contrastive Language-Image Pretraining (CLIP) model~\cite{radford2021CLIP}, pretrained on WIT comprising 400M image-text pairs collected from the internet, is utilized as the image/text encoder.
In image-text learning, the image/text encoder is trained on large-scale image-text data to embed the pairs closer in the embedding space.
The previous study VIP-ANT\cite{vipant_zhao2022g} utilizes AudioSet~\cite{Gemmeke2017Audioset}, consisting of 10-second clips collected from approximately 2M YouTube videos, for creating audio-image pairs by randomly picking images from the corresponding video streams.
This enables learning of an audio encoder that explicitly embeds audio closer to the corresponding image in the embedding space.
An improvement in the audio-text retrieval performance was then achieved with a small amount of paired audio-text data by transferring the knowledge acquired from audio-image learning.

One of the challenges of audio-image retrieval is the potential inadequacy of a single randomly selected image from a video frame to express 10 seconds of audio information.
This limitation can lead to incorrect relationships between the audio and images being learned.
As a demonstration, consider a scenario where an audio clip and images are extracted from a video that contains a large explosion towards the end. 
The most prominent feature of the audio clip is the sound of the explosion.
However, if an image is randomly selected before the sound of the explosion, where nothing significant is happening, there is a risk that the relationship will be incorrectly trained.

To tackle this issue and better match images and audio information, we investigate and evaluate two methods.
The first approach is the Nearest Match, which is the nearest neighbor matching method for image selection and relies on the distance between the video frames and the audio. 
The second approach is the Multiframe Match, in which several video frames are used simultaneously during model training, rather than selecting just one out of all video frames.
Using several frames at once is considered beneficial because it reduces the risk of selecting frames that do not match the audio information, ensuring more accurate learning of the relationship between audio and image.
As it has been shown to be beneficial to take temporal agreement into account in audio-video retrieval~\cite{morgado2021audiotexttemporalag}, audio is also divided into frames.
It is expected that performance will be improved by learning the agreement between the frames.

The rest of this paper is organized as follows. 
In Section 2, we introduce a detailed overview of multimodal learning for audio-text retrieval with images as a pivot and present our proposed methods for audio-image learning, namely Nearest Match and Multiframe Match. 
In Section 3, we detail the experiments conducted to evaluate the performance of these methods.
In Section 4, the experimental results and their implications are discussed. 
Finally, we conclude this work in Section 5.

\section{Method}
In this section, we describe three learning processes for our multimodal learning to transfer knowledge of audio-image for audio-text retrieval.
First, the model learns the co-occurrence relationship between image and text. 
Then, it extends this learning to capture the co-occurrence relationships between audio and image, as shown in Fig.~\ref{fig:multi}(i). 
Finally, the model leverages a small amount of audio-text pair data to learn the co-occurrence relationship between audio and text as shown in Fig.~\ref{fig:multi}(ii).
Through these stages, the encoders $f_I$, $f_A$, and $f_T$ are trained for image $I$, audio $A$, and text $T$, respectively. These are optimized to embed a modality pair closer in the embedding space.
This optimization is achieved through contrastive learning and the use of InfoNCE loss~\cite{sohn2016infonce}:
\begin{multline}
    \mathcal{L}_{\text{InfoNCE}}({\bf Q},{\bf K})= \\[-1pt]
    \sum_j \frac{\exp(\text{sim}({\bf q}_{j},{\bf k}_{j}))}{\sum_q \exp(\text{sim}({\bf q},{\bf k}_{j}))}+\frac{\exp(\text{sim}({\bf q}_{j},{\bf k}_{j}))}{\sum_k \exp(\text{sim}({\bf q}_{j},{\bf k}))}
\end{multline}

\vspace{2pt}
\noindent Here, ${\bf Q}$, ${\bf K}$ are two sets of data points from two different modal domains, respectively. ${\bf q}_{j}$, ${\bf k}_{j}$ are vector representations of the co-occurring pair $({\bf q}_{j},{\bf k}_{j})$ which are encoded by $f_Q({\bf q}_{j})$ and $f_K({\bf k}_{j})$.
$\text{sim}({\bf q},{\bf k})$ is the similarity between ${\bf q}$ and ${\bf k}$, which we take to be scaled cosine similarity.

\subsection{Image-text pretraining}
Instead of learning the image encoder and text encoder from scratch, we leverages pretrained CLIP model to ensure a strong alignment between image and text~\cite{radford2021CLIP}.
Specifically, we utilize the ViT-B/32 model, which consists of a 12-layer vision transformer (ViT)~\cite{dosovitskiy2020image} and a 12-layer language transformer~\cite{vaswani2017attention}.
We use encoders from the CLIP image model as $f_I$ and the text model as $f_T$.
Note that, during the learning process, they are kept frozen, ensuring that the joint image-text representation space remains unchanged.
\subsection{Audio-image pretraining (conventional)}
In this stage, the audio encoder is trained using large-scale audio and image pairs.
The architecture of the audio encoder follows that of the CLIP image encoder, using a vision transformer (ViT-B/32). 
The parameters of the audio encoder are initialized by the pretrained image encoder parameters $\theta_{I}$, the effectiveness of which has been proven in previous research~\cite{vipant_zhao2022g}.
To optimize the audio encoder parameters, the loss function expressed below is minimized.

\vspace{-3pt}
\begin{equation}
\underset{\theta_{A}}{\min}~\mathcal{L}_{\text{InfoNCE}}(A,I) 
\end{equation}
Here, $\theta_{A}$ is a learnable parameter of the audio encoder $f_A$.
The parameters of the image encoder are frozen.

The audio-image pair is sampled from a video clip. Typical approaches utilize a sparse sampling strategy~\cite{Lei2021CVPR_CLIPBERT} which extracts equally spaced video frames from a video clip. The conventional method, VIP-ANT~\cite{vipant_zhao2022g}, randomly selects an image from the set of video frames and uses the entire audio to form a pair of audio and image.

\subsection{Refined audio-image pretraining (proposal)}
We consider that a single image randomly picked may not adequately represent the entire audio clip, given that corresponding audio changes from moment to moment.
To address this problem, we propose the Nearest Match method, in which an image that better represents the corresponding audio is selected.
In addition, we propose the Multiframe Match method, in which all frames are used simultaneously instead of selecting just one frame, aiming to avoid any mismatch between audio and image information.
Fig.~\ref{fig:proposed_model} shows these methods.

\subsubsection{Nearest Match}
The similarity between audio clips and video frames is calculated using cosine distance to ensure that frames that effectively represent audio information are selected.
The frame with the highest similarity is considered to co-occur with the corresponding audio and is used in an audio-image pair for training.
However, in the early stages of training, the similarity calculation may not be mature enough. 
To mitigate this, we first randomly select a frame from the video frames for predefined epochs. 
We then move on to selecting a frame based on its similarity to the audio clip. 
The following formula represents this selection process:

\begin{equation}
\text{sim}({\bf a},{\bf i})=\begin{dcases}\underset{l\in{L}}{\max}\ \text{sim}({\bf a},{\bf i}_{l}) & \epsilon \geqq n\\ \text{sim}({\bf a},{\bf i}_{l})& \epsilon < n\end{dcases},
\end{equation}

\vspace{3pt}
\noindent where $l$ and $L$ are the index of frames and the number of frames, respectively.
${\bf a} \in \mathbb{R}^{D}$ and ${\bf i}_{l} \in \mathbb{R}^{D}$ are the audio embedding and image embedding of $l$th frame, where $D$ is the dimension of the embeddings.
$n$ denotes the epoch at which the Nearest Match begins, while $\epsilon$ represents the current number of epochs.
When $\epsilon < n$, the frames are randomly selected.

\vspace{2pt}
\subsubsection{Multiframe Match}
To avoid selecting an image that does not match the audio clip, Multiframe Match incorporates all the frames simultaneously, which is intended to reduce the likelihood of learning incorrect co-occurrence relationships.
The image encoder embeds each image, and we combine them together as ${\bf i}_\text{all} \in \mathbb{R}^{L\times D}$.
We also encode the audio features and obtain the audio encoder output with $L$ frames ${\bf a}_\text{all} \in \mathbb{R}^{L\times D}$ by the same strategy of the image encoder.
We then calculate the similarity of audio and image embeddings as follows:
\begin{equation}
    \text{sim}({\bf a},{\bf i}) = \text{sim}({\rm flatten} ( {\bf a}_\text{all} ) , {\rm flatten} ( {\bf i}_\text{all}) ).
\end{equation}

We expect that the pair of ${\bf a}_\text{all}$ and ${\bf i}_\text{all}$ enables more effective learning of correspondences between short segments and acquires explicit co-occurrence relationships between audio and image.

\subsection{Audio-text fine-tuning}
In this stage, we employ the same approach of audio-text fine-tuning~\cite{vipant_zhao2022g} that demonstrated the effectiveness of limited audio-text supervision in improving performance.
Following the audio-image pretraining stage, the audio encoder is further fine-tuned using small audio-text pairs.
This training process involves minimizing the following loss function.

\begin{equation}
\underset{\theta_{A}}{\min}~\mathcal{L}_{\text{InfoNCE}}(A,T) 
\end{equation}

\vspace{2pt}
\noindent Note that the parameters of the text encoder are frozen.

\begin{table}[t!]
\vspace{-3pt}
\centering
\caption{Experimental conditions}
\vspace{-3pt}
\label{table:conditions}
\begin{tabular}{lcc}
\toprule
\textbf{Hyperparameter} & \textbf{Audio-image} & \textbf{Audio-text} \\ \midrule
Batch size & 128 & 64 \\
Optimizer & LARS~\cite{you2017Lars} & LARS~\cite{you2017Lars} \\
Weight decay & 1.0 $\times$ $10^{-6}$ & 1.0 $\times$ $10^{-6}$  \\
LR of weight & 2.0 $\times$ $10^{-1}$ & 1.0 \\
LR of bias & 4.8 $\times$ $10^{-3}$ & 2.4 $\times$ $10^{-2}$ \\
Training epoch & 30 & 20 \\
Warmup epoch & 10 & 10 \\ 
\bottomrule
\vspace{3pt}
\end{tabular}

\centering
\caption{Number of data samples in each split}
\vspace{-3pt}
\label{table:dataset}
\begin{tabular}{lcc}
\toprule
 \textbf{Split} & \textbf{Audio-image} & \textbf{Audio-text} \vspace{1pt} \\  \midrule
Train & \begin{tabular}[c]{@{}c@{}}AudioSet (unbalanced)\\ 1,673,501\end{tabular}  \vspace{4pt} & \begin{tabular}[c]{@{}c@{}}AudioCaps \\ 43,045 (× 1 caption)\end{tabular} \vspace{4pt} \\
Val & \begin{tabular}[c]{@{}c@{}}AudioSet (balanced)\\ 16,273\end{tabular} \vspace{4pt} & \begin{tabular}[c]{@{}c@{}}AudioCaps \\  429 (× 1 caption)\end{tabular} \vspace{4pt} \\
Test & \begin{tabular}[c]{@{}c@{}}AudioSet (eval)\\ 16,000\end{tabular} \vspace{1pt} & \begin{tabular}[c]{@{}c@{}}AudioCaps \\ 845 (× 5 captions)\end{tabular} \vspace{1pt} \\
\bottomrule
\end{tabular}
\end{table}

\section{Experiments}
\begin{table*}[t!]
\small
\centering
\caption{Audio-image retrieval results}
\vspace{-1pt}
\label{tab:AI-results}
\centering
\begin{tabular}{lccccc}
\toprule
 & \multicolumn{2}{c}{A-\textgreater{}I} & \multicolumn{2}{c}{I-\textgreater{}A} \\
\cmidrule(lr){2-3} \cmidrule(lr){4-5}
Model & R@1 & R@5 & R@1 & R@5 \\
\midrule
VIP-ANT~\cite{vipant_zhao2022g} (reproduction) & $10.94\ {\pm \ 0.06}$ & $26.49\ {\pm \ 0.07}$ & $11.89\ {\pm \ 0.15}$ & $27.94\ {\pm \ 0.11}$ \\
Nearest Match (n = 0) & $10.83\ {\pm \ 0.09}$ & $26.46\ {\pm \ 0.14}$ & $11.74\ {\pm \ 0.30}$ & $27.74\ {\pm \ 0.10}$ \\
Nearest Match (n = 5) & $10.94\ {\pm \ 0.29}$ & $26.59\ {\pm \ 0.18}$ & $11.85\ {\pm \ 0.22}$ & $27.96\ {\pm \ 0.10}$ \\
Nearest Match (n = 10) & $10.88\ {\pm \ 0.18}$ & $26.58\ {\pm \ 0.21}$ & $11.95\ {\pm \ 0.08}$ & $28.01\ {\pm \ 0.27}$ \\
Nearest Match (n = 15) & $10.69\ {\pm \ 0.12}$ & $26.96\ {\pm \ 0.21}$ & $12.14\ {\pm \ 0.02}$ & $28.19\ {\pm \ 0.30}$ \\
Multiframe Match & $\mathbf{15.83\ {\pm \ 0.09}}$ & $\mathbf{33.93\ {\pm \ 0.13}}$ & $\mathbf{17.65\ {\pm \ 0.23}}$ & $\mathbf{36.27\ {\pm \ 0.19}}$ \\
\bottomrule
\end{tabular}
\end{table*}
\begin{table*}
\small
\centering
\caption{Audio-text retrieval results}
\vspace{-1pt}
\label{tab:AT-results}
\begin{tabular}{lccccc}
\toprule
 & \multicolumn{2}{c}{T-\textgreater{}A} & \multicolumn{2}{c}{A-\textgreater{}T}  \\
\cmidrule(lr){2-3} \cmidrule(lr){4-5} 
Model & R@1 & R@10 & R@1 & R@10  \\
\midrule
VIP-ANT~\cite{vipant_zhao2022g} (reproduction) & $26.17\ {\pm \ 0.10}$ & $73.85\ {\pm \ 0.94}$ & $30.85\ {\pm \ 0.78}$ & $77.36\ {\pm \ 1.26}$ \\
Nearest Match (n = 0) & $25.84\ {\pm \ 0.79}$ & $73.64\ {\pm \ 0.53}$ & $29.63\ {\pm \ 0.82}$ & $77.71\ {\pm \ 0.67}$ \\
Nearest Match (n = 5) & $26.50\ {\pm \ 0.25}$ & $73.73\ {\pm \ 0.28}$ & $31.05\ {\pm \ 0.70}$ & $77.71\ {\pm \ 0.26}$ \\
Nearest Match (n = 10) & $26.50\ {\pm \ 0.25}$ & $73.74\ {\pm \ 0.15}$ & $30.69\ {\pm \ 0.04}$ & $77.59\ {\pm \ 0.20}$ \\
Nearest Match (n = 15) & $\mathbf{26.87\ {\pm \ 0.27}}$ & $\mathbf{74.57\ {\pm \ 0.34}}$ & $\mathbf{31.80\ {\pm \ 0.66}}$ & $\mathbf{78.11\ {\pm \ 0.52}}$ \\
Multiframe Match & $25.20\ {\pm \ 0.20}$ & $71.96\ {\pm \ 0.09}$ & $29.90\ {\pm \ 0.19}$ & $75.74\ {\pm \ 0.17}$ \\
\bottomrule
\end{tabular}
\end{table*}
%
\subsection{Experimental conditions}
\subsubsection{Implementation details}
Table~\ref{table:conditions} presents model hyperparameters used in our experiments.
The audio/image encoder architecture is based on CLIP's Vision Transformer (ViT-B/32, see~\cite{radford2021CLIP} for details).
While the image ViT encodes non-overlapped image patches into a sequence of image tokens, the audio ViT taking a Mel filterbank features (FBANK) input uses the convolution stride of 16$\times$24 to allow for overlaps among neighbor patches in a convolution layer, where 16$\times$24 denotes 16-time frames and 24-frequency bins strides.
This results in about 300 audio tokens for a patch size of 32$\times$32 and an input size of 1000$\times$128.
Models are optimized by Layerwise Adaptive Rate Scaling (LARS)~\cite{you2017Lars}, with initial learning rates for model weights and model bias set to 4.8e-3 and 2e-1, respectively. 
Models were trained on an NVIDIA RTX A6000 GPU (48G) to learn audio and image relationships between for 30 epochs.
The model was then trained on 20 epochs of audio and text relationships.
The batch size was set to 128 for audio-image pretraining and 64 for audio-text fine-tuning to match the GPU performance.
The experimental results are the average of the evaluation results using three random seeds for initializing model parameters.
Kaldi~\cite{Povey_ASRU2011kaldi} is used to create FBANK features from raw audio signals with a Hanning window, 128 triangular Mel frequency bins, and a 10 ms frame shift.
If an audio clip has multiple channels, the first audio channel is always used. 
Before applying Kaldi, we normalize the raw audio clips by subtracting the mean from them. 
Additionally, we compute the mean and standard deviation of FBANK features on the AudioSet training set. 
Then, we normalize the FBANK features of each audio clip using these computed statistics.
The other experimental conditions follow the conventional VIP-ANT experiment settings.
%
%
\subsubsection{Dataset}
AudioSet~\cite{Gemmeke2017Audioset} is used for audio-image pretraining.
AudioSet consists of an extended ontology of 527 audio event classes and a collection of about 2M human-labeled audio clips of approximately 10 seconds extracted from YouTube videos.
AudioSet provides balanced and unbalanced training sets.
The balanced set contains 22K audio clips, while the unbalanced set contains approximately 2M audio clips.
We also used AudioCaps~\cite{kim-NAACL-HLT-2019AudioCaps}, which humans annotated via the Amazon Mechanical Turk (AMT) crowdsource platform for approximately 10 seconds of audio clips extracted from YouTube videos, for audio-text fine-tuning.
Each audio clip in the training set has one human-annotated caption, while each clip in the validation and test sets has five ground-truth captions.
The number of image frames $L$ is set to four.
Image frames are extracted every 3.3 seconds approximately from a 10-second video in AudioSet.
Note that many video clips in both AudioSet and AudioCaps are no longer available on YouTube, and we used about 82\% and 87\% of each dataset from the original list, as described in Table \ref{table:dataset}.

\subsubsection{Evaluation metrics}
Audio-image pretraining and audio-text fine-tuning are evaluated using R@k (Recall at rank k), a standard retrieval metric.
R@k represents the proportion of targets retrieved from the top k results, with higher values indicating better retrieval performance.
\section{Results and Discussion}
\label{sec:illust}
Tables~\ref{tab:AI-results} and \ref{tab:AT-results} show the experimental results of audio-image cross retrieval (A$\rightarrow$I, I$\rightarrow$A) and audio-text cross retrieval (A$\rightarrow$T, T$\rightarrow$A) in terms of R@k with standard deviations.
Here, $n$ denotes the epoch at which the Nearest Match begins.
Note that VIP-ANT (reproduction) is a conventional model~\cite{vipant_zhao2022g} reproduced in our environment since it is not possible to make an exact comparison owing to the dataset size\footnote{About 18\% of AudioSet samples are no longer available, making it difficult to compare ours with the previous papers directly.} and the batch size because of a different GPU performance.

Table~\ref{tab:AI-results} shows that the Multiframe Match method significantly improves the performance of audio-image retrieval compared to the conventional approach of random frame selection.
On the other hand, the Nearest Match method showed only a slight performance improvement.
This suggests that since audio and video frames are aligned to some extent, it is better for audio-image retrieval to train the temporal relationship between the audio and video frames by utilizing a series of frames, rather than selecting a single prominent frame.

Table~\ref{tab:AT-results} shows that the Nearest Match method outperforms the conventional random selection method.
This is because audio captions tend to be annotated to dominant sounds in a sound clip, and Nearest Match, which selects a prominent frame that is highly similar to the sound, may have worked more effectively than Multiframe Match.
In addition, Nearest Match ($n=15$) shows the highest performance in A$\rightarrow$T and T$\rightarrow$A than when ($n=0,5,10$).
This shows that waiting for the audio encoder to mature and starting intensive learning of the most similar audio-image pairs is effective in bringing audio closer to text in CLIP's embedding space.
%
\section{Conclusion}
\label{sec:page}
In this study, we proposed two methods to refine knowledge transfer on audio-image temporal agreement for audio-text cross retrieval.
Conventional approaches in audio-image learning assign a single image randomly picked from the corresponding video stream to the entire audio clip, assuming their co-occurrence. 
As conventional approaches may not accurately capture the temporal agreement between the target audio and image, we proposed two methods for audio and image assignment that effectively capture the temporal information.
Experimental results demonstrated that Nearest Match could improve the A$\rightarrow$T performance by selecting the most similar image to the audio, particularly during the later epochs of the training process.
In contrast, Multiframe Match improved the performance of A$\rightarrow$I.
These results showed that refining audio-image temporal agreement contributes to the transfer of better knowledge to audio-text retrieval.
One of the future tasks is to analyze the amount of data in the dataset where mismatches between audio and image can occur in detail and to analyze the data for which the two proposed methods were effective.

\bibliographystyle{IEEEtran}
\bibliography{refs}

\begin{thebibliography}{10}
\providecommand{\url}[1]{#1}
\csname url@samestyle\endcsname
\providecommand{\newblock}{\relax}
\providecommand{\bibinfo}[2]{#2}
\providecommand{\BIBentrySTDinterwordspacing}{\spaceskip=0pt\relax}
\providecommand{\BIBentryALTinterwordstretchfactor}{4}
\providecommand{\BIBentryALTinterwordspacing}{\spaceskip=\fontdimen2\font plus
\BIBentryALTinterwordstretchfactor\fontdimen3\font minus
  \fontdimen4\font\relax}
\providecommand{\BIBforeignlanguage}[2]{{%
\expandafter\ifx\csname l@#1\endcsname\relax
\typeout{** WARNING: IEEEtran.bst: No hyphenation pattern has been}%
\typeout{** loaded for the language `#1'. Using the pattern for}%
\typeout{** the default language instead.}%
\else
\language=\csname l@#1\endcsname
\fi
#2}}
\providecommand{\BIBdecl}{\relax}
\BIBdecl

\bibitem{MulRet_Kaiye2016}
K.~Wang, Q.~Yin, W.~Wang, S.~Wu, and L.~Wang, ``A comprehensive survey on
  cross-modal retrieval,'' \emph{arXiv preprint arXiv:1607.06215}, 2016.

\bibitem{Vonam2019Imagetext}
N.~Vo, L.~Jiang, C.~Sun, K.~Murphy, L.-J. Li, L.~Fei-Fei, and J.~Hays,
  ``Composing text and image for image retrieval - an empirical odyssey,'' in
  \emph{IEEE/CVF Conf. Comput. Vis. Pattern Recognit. (CVPR)}, 2019, pp.
  6432--6441.

\bibitem{cao2022itsurvey}
M.~Cao, S.~Li, J.~Li, L.~Nie, and M.~Zhang, ``Image-text retrieval: A survey on
  recent research and development,'' in \emph{Int. Jt. Conf. Artif. Intell.},
  2022.

\bibitem{fang2021clip2video}
H.~Fang, P.~Xiong, L.~Xu, and Y.~Chen, ``{CLIP2Video}: Mastering video-text
  retrieval via image {CLIP},'' \emph{arXiv preprint arXiv:2106.11097}, 2021.

\bibitem{xu2021videoclip}
H.~Xu, G.~Ghosh, P.-Y. Huang, D.~Okhonko, A.~Aghajanyan, F.~Metze,
  L.~Zettlemoyer, and C.~Feichtenhofer, ``{VideoCLIP}: Contrastive pre-training
  for zero-shot video-text understanding,'' in \emph{Conf. Empir. Methods Nat.
  Lang. Process. (EMNLP)}, 2021, pp. 6787--6800.

\bibitem{guzhov2022audioclip}
A.~Guzhov, F.~Raue, J.~Hees, and A.~Dengel, ``{AudioCLIP}: Extending clip to
  image, text and audio,'' in \emph{IEEE Int. Conf. Acoust. Speech Signal
  Process. (ICASSP)}.\hskip 1em plus 0.5em minus 0.4em\relax IEEE, 2022, pp.
  976--980.

\bibitem{wu2022wav2clip}
H.-H. Wu, P.~Seetharaman, K.~Kumar, and J.~P. Bello, ``{Wav2CLIP}: Learning
  robust audio representations from clip,'' in \emph{IEEE Int. Conf. Acoust.
  Speech Signal Process. (ICASSP)}.\hskip 1em plus 0.5em minus 0.4em\relax
  IEEE, 2022, pp. 4563--4567.

\bibitem{ruan2023accommodating}
L.~Ruan, A.~Hu, Y.~Song, L.~Zhang, S.~Zheng, and Q.~Jin, ``Accommodating audio
  modality in {CLIP} for multimodal processing,'' \emph{arXiv preprint
  arXiv:2303.06591}, 2023.

\bibitem{akbari2021vatt}
H.~Akbari, L.~Yuan, R.~Qian, W.-H. Chuang, S.-F. Chang, Y.~Cui, and B.~Gong,
  ``{VATT}: Transformers for multimodal self-supervised learning from raw
  video, audio and text,'' \emph{Adv. Neural Inf. Process. Syst (NeurIPS)},
  vol.~34, pp. 24\,206--24\,221, 2021.

\bibitem{TexAud_Oncescu2021}
A.-M. Oncescu, A.~S. Koepke, J.~Henriques, Z.~Akata, and S.~Albanie, ``Audio
  retrieval with natural language queries,'' in \emph{INTERSPEECH}, 2021.

\bibitem{Srinivasan2021WIT}
K.~Srinivasan, K.~Raman, J.~Chen, M.~Bendersky, and M.~Najork, ``{WIT}:
  Wikipedia-based image text dataset for multimodal multilingual machine
  learning,'' in \emph{Int. ACM SIGIR Conf. Res. Dev. Inf. Retr.}\hskip 1em
  plus 0.5em minus 0.4em\relax {ACM}, jul 2021.

\bibitem{chen2015cococap}
X.~Chen, H.~Fang, T.-Y. Lin, R.~Vedantam, S.~Gupta, P.~Doll{\'a}r, and C.~L.
  Zitnick, ``Microsoft {COCO} captions: Data collection and evaluation
  server,'' \emph{arXiv preprint arXiv:1504.00325}, 2015.

\bibitem{drossos2020clotho}
K.~Drossos, S.~Lipping, and T.~Virtanen, ``Clotho: An audio captioning
  dataset,'' in \emph{IEEE Int. Conf. Acoust. Speech Signal Process.
  (ICASSP)}.\hskip 1em plus 0.5em minus 0.4em\relax IEEE, 2020, pp. 736--740.

\bibitem{kim-NAACL-HLT-2019AudioCaps}
C.~D. Kim, B.~Kim, H.~Lee, and G.~Kim, ``{{AudioCaps}: Generating Captions for
  Audios in The Wild},'' in \emph{N. Am. Chapter Assoc. Comput. Linguist.: Hum.
  Lang. Technol. (NAACL-HLT)}, 2019.

\bibitem{Koepke2022atretrievalque}
A.~S. Koepke, A.-M. Oncescu, J.~Henriques, Z.~Akata, and S.~Albanie, ``Audio
  retrieval with natural language queries: A benchmark study,'' \emph{IEEE
  Trans. Multimed.}, pp. 1--1, 2022.

\bibitem{vipant_zhao2022g}
Y.~Zhao, J.~Hessel, Y.~Yu, X.~Lu, R.~Zellers, and Y.~Choi, ``Connecting the
  dots between audio and text without parallel data through visual knowledge
  transfer,'' in \emph{N. Am. Chapter Assoc. Comput. Linguist. (NAACL)}, Jul.
  2022, pp. 4492--4507.

\bibitem{radford2021CLIP}
A.~Radford, J.~W. Kim, C.~Hallacy, A.~Ramesh, G.~Goh, S.~Agarwal, G.~Sastry,
  A.~Askell, P.~Mishkin, J.~Clark \emph{et~al.}, ``Learning transferable visual
  models from natural language supervision,'' in \emph{Int. Conf. Mach.
  Learn.}, 2021, pp. 8748--8763.

\bibitem{Gemmeke2017Audioset}
J.~F. Gemmeke, D.~P.~W. Ellis, D.~Freedman, A.~Jansen, W.~Lawrence, R.~C.
  Moore, M.~Plakal, and M.~Ritter, ``{Audio Set}: An ontology and human-labeled
  dataset for audio events,'' in \emph{IEEE Int. Conf. Acoust. Speech Signal
  Process. (ICASSP)}, 2017, pp. 776--780.

\bibitem{morgado2021audiotexttemporalag}
P.~Morgado, N.~Vasconcelos, and I.~Misra, ``Audio-visual instance
  discrimination with cross-modal agreement,'' in \emph{IEEE/CVF Conf. Comput.
  Vis. Pattern Recognit. (CVPR)}, 2021, pp. 12\,475--12\,486.

\bibitem{sohn2016infonce}
K.~Sohn, ``Improved deep metric learning with multi-class {N}-pair loss
  objective,'' \emph{Adv. Neural Inf. Process. Syst.}, vol.~29, 2016.

\bibitem{dosovitskiy2020image}
A.~Dosovitskiy, L.~Beyer, A.~Kolesnikov, D.~Weissenborn, X.~Zhai,
  T.~Unterthiner, M.~Dehghani, M.~Minderer, G.~Heigold, S.~Gelly \emph{et~al.},
  ``An image is worth 16x16 words: Transformers for image recognition at
  scale,'' \emph{arXiv preprint arXiv:2010.11929}, 2020.

\bibitem{vaswani2017attention}
A.~Vaswani, N.~Shazeer, N.~Parmar, J.~Uszkoreit, L.~Jones, A.~N. Gomez,
  {\L}.~Kaiser, and I.~Polosukhin, ``Attention is all you need,'' \emph{Adv.
  Neural Inf. Process. Syst. (NeurIPS)}, vol.~30, 2017.

\bibitem{Lei2021CVPR_CLIPBERT}
J.~Lei, L.~Li, L.~Zhou, Z.~Gan, T.~L. Berg, M.~Bansal, and J.~Liu, ``Less is
  more: {ClipBERT} for video-and-language learning via sparse sampling,'' in
  \emph{IEEE/CVF Conf. Comput. Vis. Pattern Recognit. (CVPR)}, June 2021, pp.
  7331--7341.

\bibitem{you2017Lars}
Y.~You, I.~Gitman, and B.~Ginsburg, ``Scaling {SGD} batch size to 32k for
  imagenet training,'' \emph{arXiv preprint arXiv:1708.03888}, 2017.

\bibitem{Povey_ASRU2011kaldi}
D.~Povey, A.~Ghoshal, G.~Boulianne, L.~Burget, O.~Glembek, N.~Goel,
  M.~Hannemann, P.~Motlicek, Y.~Qian, P.~Schwarz, J.~Silovsky, G.~Stemmer, and
  K.~Vesely, ``The kaldi speech recognition toolkit,'' in \emph{IEEE Workshop
  Autom. Speech Recognit. Underst. (ASRU)}, Dec. 2011.

\end{thebibliography}

\end{document}